\documentclass[a4paper,12pt,onecolumn,oneside,iopams,setstack,amsmath,amssymb,amsfonts]{iopart}

\usepackage[pdftex]{hyperref}                    % create links and bookmarks in the pdf file
\usepackage[pdftex]{graphicx}
\usepackage{cite}
\usepackage{dcolumn}                             % Align table columns on decimal point
\usepackage{bm}                                  % bold math
\usepackage[verbose,tmargin=3cm,bmargin=3cm,lmargin=3cm,rmargin=3cm]{geometry}
\usepackage[switch,running]{lineno}              % line number package

\newcommand{\dif}{\mathrm{d}}                    % differential
\newcommand{\me}{\mathrm{e}}                     % math e
\begin{document}

%\begin{linenumbers}

\title[Multiple scattering representation for radiation trapping]
      {\Large
       Advantages of the multiple scattering representation approach to radiation trapping}

\author{E~J~Nunes-Pereira$^{1, \, \dagger}$, A~R~Alves-Pereira$^1$,
  J~M~G~Martinho$^2$ and M~N~Berberan-Santos$^2$}

\address{$^1$ Universidade do Minho, Escola de Ci\^{e}ncias,
  Centro de F\'{i}sica, 4710-057 Braga, Portugal}

\address{$^2$ Centro de Qu\'{i}mica-F\'{i}sica Molecular, Instituto
  Superior T\'{e}cnico, 1049-001 Lisboa, Portugal}

\ead{$^{\dagger}$~epereira@fisica.uminho.pt}

\date{\today}

\begin{abstract}

A simple stochastic formulation of the multiple scattering representation solution of the classical
linear incoherent trapping problem is presented for a broad audience. A clear connection with the
alternative Holstein's solution ansatz is emphasized by the (re)interpretation of the fundamental
mode as the one associated with a relaxed nonchanging spatial distribution of excitation.
Expressions for overall relaxation parameters~(ensemble emission yield and lifetime) as well as
time-resolved~(decay and spatial distribution) and steady-state quantities~(spectra and spatial
distribution) are given with the fundamental mode contribution singled out. The multiple scattering
representation is advocated for final undergraduate and beginning graduate physics instruction
based on physical insight and computation feasibility. This will be illustrated in the following
instalment of this contribution.

\end{abstract}

\maketitle

%%%%%%%%%%%%%%%%%%%%%%%%%%%%%%%%%%%%%%%%%%%%%%%%%%%%%%%%%%%%%%%%%%%%%%%%%%%%%%%%
%%% Introduction
%%%%%%%%%%%%%%%%%%%%%%%%%%%%%%%%%%%%%%%%%%%%%%%%%%%%%%%%%%%%%%%%%%%%%%%%%%%%%%%%

%
\section{Introduction}\label{Introduction}

Radiation trapping is important in areas as diverse as stellar atmospheres~\cite{Mihalas1978:book},
plasmas and atomic vapors luminescence~\cite{Molisch1998:book}, terrestrial atmosphere and ocean
optics~\cite{Thomas1999:book}, molecular luminescence~\cite{Berberan-Santos1999:chapter}, infrared
radiative transfer~\cite{Modest2003:book} and cold atoms~\cite{Bardou2002:book}. In optically thick
media, the emitted resonant radiation suffers several reabsorption and reemission events before
eventually escaping to the exterior; the radiation is said to be imprisoned or trapped. Atomic
\textit{radiation trapping} is also known as \textit{imprisonment of resonance radiation},
\textit{reabsorption}, \textit{self-absorption}, \textit{line transfer}, \textit{radiation
diffusion} or \textit{multiple scattering} of resonance radiation.

The study of electronic excitation energy trapping has played a central role in the study of
collective excitation relaxation in atomic physics during the whole of the last century. After the
first experimental studies with mercury vapours in the 1910s by Robert Wood, Theodore Holstein,
working on atomic physics at Westinghouse Research Laboratories between 1941 and 1959, driven by
the need to quantify trapping in fluorescence lamps, laid down in 1947 a general framework to deal
with incoherent trapping~\cite{Holstein1947:PhysRev.72.1212}. Independently, Leon Biberman, in
Russia~(then Soviet Union), derived the same key equation for radiation transfer in spectral
lines~\cite{Biberman1947:ZhFiz.17.416}. This equation, to be discussed below, is now known as the
Holstein-Biberman equation and occupies a central place in the study of the kinetics and transport
of excited resonance states in gases and plasmas and in condensed media. During the following
decades, the forefront of trapping studies was conducted in an astrophysical context until it was
realized in the mid 1990s that trapping posed several fundamental limitations to the cooling of
atomic ensembles. Recently it was shown that incoherent atomic trapping is probably the simplest
and best characterized case of a L\'{e}vy flight found up to
now~\cite{Nunes-Pereira2004:PhysRevLett.93.120201} and this is important in the recent view that
anomalous diffusion should be treated in a similar footing with normal, Brownian-type,
diffusion~\cite{Metzler-Anomalous-2000:PhysRep.399.1&Metzler-Anomalous-2004:JPhysA.37.R161&SuperdiffusionI:JCP.125.174308}.
Or, stated otherwise, that the generalized forms of the well known Central Limit Theorem are much
more important in practical situations than once generally thought. From the point of view of
economically important applications, electric discharge lamps are still the most important
application~\cite{Lister2004:RMP.76.541&Proud1986:book&Waymouth1971:book} eventhough electrodeless
fluorescence lamps~\cite{Kushner2004:JPhysD.37.R161} and large area plasma display
panels~\cite{KushnerPDPsB:JApplPhys.85.3460&KushnerPDPsC:JApplPhys.85.3470&KushnerPDPsA:JApplPhys.87.2700}
are gaining increasing importance.

In spite of its importance from the fundamental physics point of view as well as from the need to
control trapping in practical applications, the discussion of the physical implications of
radiation trapping at the level of the nonspecialist physics major has been hampered by several
factors, most notably, relying on the standard use of Holstein original mode expansion and the
subsequent intricate computational technicalities involved in obtaining useful estimates of
trapping dependent quantities. In this context, the present contribution should appeal to a broad
audience and be valuable in the context of final undergraduate and beginning graduate physics
instruction. The emphasis can be tailored either more towards a mathematical~(this paper) or,
alternatively, a computational physics project~(following instalment), always with physical insight
as the primary goal.

The Multiple Scattering Representation~(MSR) solution for the trapping dynamics in linear
incoherent conditions is given here a sound stochastic formulation and a clear connection with the
older Holstein ansatz emphasized. The dynamics of incoherent trapping are first discussed
in~\sref{Dynamics of incoherent trapping} and expressions for ensemble overall relaxation
parameters as well as spatial distributions are derived. These are extended for steady-state
observables under linear system response conditions in~\sref{Steady-state conditions}.
In~\sref{Fundamental mode}, the emphasis is put on the interpretation of the so-called Holstein
fundamental mode within the~MSR general framework. Finally,~\sref{Conclusions} presents the main
conclusions. The evaluation of trapping dependent quantities is postponed to the following
instalment of this work, with focus in the computational advantages of the proposed model for
reabsorption.

%%%%%%%%%%%%%%%%%%%%%%%%%%%%%%%%%%%%%%%%%%%%%%%%%%%%%%%%%%%%%%%%%%%%%%%%%%%%%%%%
%%% Dynamics of incoherent trapping
%%%%%%%%%%%%%%%%%%%%%%%%%%%%%%%%%%%%%%%%%%%%%%%%%%%%%%%%%%%%%%%%%%%%%%%%%%%%%%%%

%
\section{Dynamics of incoherent trapping}\label{Dynamics of incoherent trapping}

Any description of radiation trapping must take into account the non-local character of
reabsorption on the excitation dynamics of an ensemble of excited species. Non-local effects pose
several difficulties to obtain a solution for the ensemble dynamics, most notably, (i)~local
excitation occurs by reabsorption of radiation originating from any point within the sample~(high
computational demands since the actual geometry should be fully included in the mathematical
description), (ii)~the need to be able to describe the time between emission and subsequent
reabsorption in a different coordinate~(the spatial and temporal evolutions become convoluted) and
(iii)~non-linear effects~(partial saturation of absorption and stimulated emission rendering the
whole of the dynamics dependent in a complicated way upon the local radiation densities). However,
for the classical trapping problem~\cite{Molisch1998:book}, in which the time-of-flight of
in-transit radiation between emission and reabsorption is negligible compared with the natural
lifetime of the excited states, and for linear response conditions, some approximations are
possible that allow a formidable simplification of the solution of the trapping problem.

\subsection{Characteristic scales}\label{Characteristic scales}

A considerable simplification is achieved by using characteristic \textit{time} and \textit{length}
scales for trapping: the temporal and spatial dynamics become dimensionless in a scaled time and
dimensionless \textit{optical density} or \textit{opacity} distance. The scaled time is just
$t=\Gamma t^{\prime }$, where $\Gamma $ is the global excitation deactivation~(radiative plus
non-radiative) rate constant. The scaled distance however should reflect both the whole of the
absorption spectrum as well as the dimension and density of the ensemble. One should use for the
absorption lineshape~$\Phi \left( x\right)$ a normalized probability distribution function~(so
that~$\int_{-\infty }^{+\infty }\Phi \left( x\right) \,\dif x = 1$; note that~$x$ was chosen as a
scaled difference to the center of line frequency). With this, a monochromatic line opacity along a
given pathlength~$l$, for homogeneously distributed species, can be defined as $k\left( x\right)
=n\sigma _{0} l \Phi \left( x\right) / \Phi \left( 0\right) = k_{0}\Phi \left( x\right) / \Phi
\left( 0\right) = \Phi \left( x\right) r$, where~$\Phi \left( 0\right)$ and~$\sigma _{0}$ are the
center-of-line normalized absorption coefficient and cross section. The monochromatic opacity is
proportional to the number density~$n$ and to the center-of-line opacity~$k_{0} =n\sigma _{0} l$.
Finally, the overall~(reflecting the whole of the spectral distribution) dimensionless opacity
is~$r=\int_{-\infty }^{+\infty }k\left( x\right) \,\dif x = \frac{k_{0}}{\Phi \left( 0\right)}$ and
this defines the characteristic lengthscale for trapping.

\subsection{Holstein-Biberman equation}\label{Holstein-Biberman equation}

The starting point of the majority of incoherent trapping models is the so-called Holstein-Biberman
equation, a Boltzman-type integro-differential equation, that gives the spatial and temporal
evolution of the excited state number density~$n(\bi{r},t)$ as

\begin{equation}
  \label{Holstein-Equation}
  \frac{\partial n\left( \bi{r},t\right)}{\partial t}=-\Gamma n\left( \bi{r},t\right) +\Gamma
  \phi_{0}\int_{V}\! f\left( \bi{r,r}^{\prime }\right) n\left(\bi{r}^{\prime },t\right)
  \dif \bi{r}^{\prime } \mbox{,}
\end{equation}

where the non-local character of trapping is evident in the last term of the right hand side: local
excited number density increase due to reabsorption of radiation emitted in all of the sample
enclosure. $\phi _{0}$ is the intrinsic, trapping undistorted radiative emission quantum yield, and
can be interpreted as the probability of photon emission by an excited state~(the ratio of
radiative over global relaxation rate constants; $\phi _{0}=\frac{\Gamma _{r}}{\Gamma _{r}+\Gamma
_{nr}}$). $f\left( \bi{r},\bi{r}^{\prime }\right)$ is the (conditional) transition probability of
photon absorption at~$\bi{r}$, given that there was emission at~$\bi{r}^{\prime }$. In this form,
Holstein-Biberman equation neglects the time-of-flight of radiation and therefore the spatial and
temporal dynamics are decoupled. Holstein proposed an eigenmode expansion

\begin{equation}
  \label{Holstein-Solution}
  n\left( \bi{r},t\right) =\sum_{n}n_{n}\left( \bi{r}\right) \me ^{-\beta _{n}t} \mbox{,}
\end{equation}

as the general solution of~\eref{Holstein-Equation}. This solution has however several important
shortcomings:~(i)~the eigenmodes~(stationary spatial modes) $n_{n}\left( \bi{r}\right)$ have a
troublesome physical interpretation since all but the slowest decaying mode have negative
values~(individual modes cannot be identified with physical distributions); (ii)~individual
relaxation constants~$\beta _{n}$ have no simple connection with physical parameters; (iii)~the
eigenmodes/values are not easily estimated; (iv)~it is very difficult to generalize the mode
expansion to additional effects~(partial frequency redistribution between absorption and
reemission, radiation propagation time, particle diffusion); and (v)~it cannot be used to obtain
the polarization of emitted radiation. However, an alternative mode expansion exists that overcomes
most of these difficulties. It is known as the Multiple Scattering Representation~(MSR) since it
identifies each spatial mode with the spatial distribution of excited species after several
scattering~(reemission-reabsorption) orders. These multiple scattering modes are associated with
several \textit{generations} of excited species, paralleling the members of decaying radioactive
families. The~MSR was independently proposed for atomic~\cite{Falecki1989:ZPhysD.14.111} and
molecular trapping~\cite{Martinho1989:JCP.90.53} and subsequently proven to be equivalent to the
original Holstein solution~\cite{Lai1993:ZPhysD.27.223&Lai1993:OpticsComm.99.316}. For general
reviews see~\cite{Molisch1998:book,Berberan-Santos1999:chapter}.

The~MSR approach is amenable to a straightforward stochastic formulation, as the sought for
quantities are the spatial and temporal probability distributions of the above mentioned
\textit{generations}. The overall dynamics must of course reflect the contribution of each
generation to the whole, and this can be obtained from each generation population
efficiency~$a_{n}$~(probability of primordial excitation populating the~$n^{th}$ generation due to
$n-1$ scattering events in cell with no escape; the first generation of excited species is formed
by the set of excited species initially created by processes other than reabsorption, the second
generation ones are the species created by trapping of first generation radiation, and so forth).
The~MSR solution ansatz to the Holtein-Biberman equation is therefore

\begin{equation}
  \label{MSRSol}
  n\left( \bi{r},t\right) =\sum_{n} a_{n} \, p_{n}\left( \bi{r}\right) g_{n}\left( t\right)
  \mbox{,}
\end{equation}

where the spatial and temporal trapping specific relaxations are given by the (normalized)
spatial~$p_{n}\left( \bi{r}\right)$ and temporal~$g_{n}\left( t\right)$ distributions. The
$p_{n}\left( \bi{r}\right)$'s substitute for the eigenmodes of the Holstein ansatz with distinctive
advantages since their physical interpretation is clear; they are the spatial distribution of
excitation after $n-1$ scattering events. The temporal distributions are easily obtained for
incoherent trapping; the temporal evolution of each generation, $g_{n}\left( t\right) $, is a
iterated convolution of the intrinsic response of each generation, $g\left( t\right) =\me ^{-t}$.
The temporal evolution is therefore~$g_{n}\left( t\right) =\frac{t^{n-1}}{\left( n-1\right) !}
\me^{-t}$~\cite{Berberan-Santos1999:chapter}. The trapping efficiency can be discussed either based
on each generation population efficiency~$a_{n}$ or, preferably, on each generation reabsorption
probability defined as~$\alpha _{n}\equiv \frac{a_{n+1}}{a_{n}}$. One should elaborate a little
further by factoring out trapping specific effects~(opacity scales, geometry, spectral
distribution) from the trivial influence of the reemission probability~$\phi _{0}$. This decoupling
can be made  by writing~$\alpha _{n}\equiv \alpha _{n}^{T}\phi _{0}$ (superscript $T$ signals
quantities dependent only on trapping efficiencies) and $a_{n}=a_{n}^{T} \, \phi _{0}^{n-1}$
with~$a_{n}^{T}=\prod\limits_{n=1}^{n-1}\alpha _{n}^{T}$. Finally, one should be aware that not
whole of the radiation emitted by each generation escapes~(or, is reabsorbed); the generation
dependent mean escape probability can be defined as~$q_{n}=\phi _{0}\left( 1-\alpha
_{n}^{T}\right)$. With these aspects in mind, useful expressions can be derived for the relevant
parameters of the ensemble.

\subsection{Ensemble relaxation}\label{Ensemble relaxation}

The trapping dynamics reflects itself in the two single most important macroscopic parameters for
the ensemble relaxation; the overall reemission efficiency~$\phi $ (mean photon reemission
probability out of sample enclosure, irrespective of scattering order) and mean excitation
deactivation or photon emission lifetime~$\tau $ (mean excitation survival time in macroscopic
ensemble). These will be the most important quantities to be extracted from~\eref{MSRSol}. Before
showing explicitly how these quantities are obtained, let us consider the relaxation from an
initially created population of excited species based on trapping alone. After a sufficiently high
number of scattering events, initial excitation will relax to a distribution which, when
normalized, does not change any more since each point deactivation is exactly balanced with local
reabsorption due to emission from the whole ensemble. Thus, based on physical insight, in all of
the following we can divide the contribution of all the generations in two groups: one summing up
the spatial changing excited species and the other grouping all the generations with a nonchanging
distribution~(with an analytical explicit sum; see discussion below). Generations will be grouped
into up to~$m=n_{nc}$ and~$m$ onwards, where the subscript is a remainder for \textit{nonchanging}.
The nonchanging distribution is stationary in the sense that is time independent but we will keep
the~$nc$ subscript to emphasize the difference to the steady-state or \textit{stationary} system
response to a continuous perturbation and avoid common misinterpretations. Two needed but less
immediate math notes are the equalities~$\sum_{n=1}^{m}nr^{n}=\frac{r}{\left( 1-r\right)
^{2}}\left[ mr^{m+1}-\left( m+1\right) r^{m}+1\right]$ and~$\sum_{n=0}^{m}\frac{x^{n}}{n!}=\me
^{x}\frac{\Gamma \left( m+1,x\right) }{m!}$, with~$\Gamma \left( a,z\right)$ the incomplete Gamma
function, both valid when~$\left\vert r\right\vert <1$.

To obtain the ensemble dynamics we will need to compute sums of the type~$\sum q_{n}a_{n}$, $\sum
nq_{n}a_{n}$ and $\sum a_{n}p_{n}g_{n}\left( t\right)$ and, in these, we will use the fact
that~$q_{n\geq m}=q_{nc}$, $p_{n\geq m}\left( \bi{r} \right) =p_{nc}\left( \bi{r} \right)$ and
$a_{n\geq m}=a_{nc}\alpha _{nc}^{n-m}$. The ensemble reemission yield is

\begin{eqnarray}
\label{phi} \phi
   & = & \int_{0}^{+\infty }\left[ \sum_{n=1}^{+\infty
}q_{n}a_{n}g_{n}\left( t\right) \right] \dif t             \nonumber \\
   & = & \sum_{n=1}^{+\infty
}q_{n}a_{n}\int_{0}^{+\infty }g_{n}\left( t\right) \dif t  \nonumber \\
   & = & \sum_{n=1}^{+\infty }q_{n}a_{n}                   \nonumber \\
   & = & \sum_{n=1}^{m-1}q_{n}a_{n}+\frac{q_{nc}a_{nc}}{\alpha
_{nc}^{m}}\sum_{n=m}^{+\infty }\alpha _{nc}^{n}            \nonumber \\
   & = & \sum_{n=1}^{m-1}q_{n}a_{n}+\frac{q_{nc}a_{nc}}{\alpha
_{nc}^{m}}\left[ \sum_{n=1}^{+\infty }\alpha
_{nc}^{n}-\sum_{n=1}^{m-1}\alpha _{nc}^{n}\right]          \nonumber \\
   & = & \sum_{n=1}^{m-1}q_{n}a_{n}+\frac{q_{nc}a_{nc}}{\alpha
_{nc}^{m}}\left[ \frac{\alpha _{nc}}{1-\alpha _{nc}}-\alpha
_{nc}\frac{1-\alpha _{nc}^{m-1}}{1-\alpha _{nc}}\right]    \nonumber \\
  & = & \sum_{n=1}^{m-1}q_{n}a_{n}+\frac{q_{nc}a_{nc}}
  {1-\alpha_{nc}}
  \mbox{,}
\end{eqnarray}

since

\begin{equation}
  \label{decay}
  \rho \left( t\right) =\sum_{n=1}^{+\infty }q_{n}a_{n}g_{n}\left( t\right)
  \mbox{,}
\end{equation}

is just the emission decay.

The mean lifetime is

\begin{equation}
  \label{lifetime}
  \tau =\frac{\int_{0}^{+\infty }t\rho \left( t\right) \dif t}{\int_{0}^{+\infty }\rho \left(
t\right) \dif t}
  \mbox{.}
\end{equation}

The denominator is just the reemission yield, while the numerator can be written as

\begin{eqnarray}
\int_{0}^{+\infty }t\left[ \sum_{n=1}^{+\infty
}q_{n}a_{n}g_{n}\left( t\right) \right] \dif t =           \nonumber \\
\quad = \sum_{n=1}^{+\infty }q_{n}a_{n}
\int_{0}^{+\infty }tg_{n}\left( t\right) \dif t            \nonumber \\
\quad = \sum_{n=1}^{+\infty }nq_{n}a_{n}                     \nonumber \\
\quad = \sum_{n=1}^{m-1}nq_{n}a_{n}+\frac{q_{nc}a_{nc}}{\alpha _{nc}^{m}}\left[ \sum_{n=1}^{+\infty
}n\alpha
_{nc}^{n}-\sum_{n=1}^{m-1}n\alpha _{nc}^{n}\right]         \nonumber \\
\quad = \sum_{n=1}^{m-1}nq_{n}a_{n}+                         \nonumber \\
\qquad + \frac{q_{nc}a_{nc}}{\alpha _{nc}^{m}}\left[ \frac{\alpha _{nc}}{\left( 1-\alpha
_{nc}\right) ^{2}}- \frac{\alpha _{nc}}{\left( 1-\alpha _{nc}\right) ^{2}}\left( \left( m-1\right)
\alpha
_{nc}^{m}-m\alpha _{nc}^{m-1}-1\right) \right]             \nonumber \\
\quad = \sum_{n=1}^{m-1}nq_{n}a_{n}+\frac{q_{nc}a_{nc}}{\left( 1-\alpha _{nc}\right) ^{2}}\left[
m\left( 1-\alpha _{nc}\right) +\alpha _{nc}\right]
  \mbox{,}
\end{eqnarray}

since each generation will decay on average in~$n$ units of the dimensionless time. Finally, the
scaled lifetime is

\begin{eqnarray}
  \label{Tau}
  \tau =\frac{\sum_{n=1}^{m-1}nq_{n}a_{n}+
  \frac{q_{nc}a_{nc}}{\left(1-\alpha _{nc}\right) ^{2}}\left[ m\left( 1-\alpha _{nc}\right)
  +\alpha _{nc}\right] }{\phi }
  \mbox{.}
\end{eqnarray}

\subsection{Spatial excitation distribution}\label{Spatial excitation distribution}

The overall time resolved normalized spatial distribution is given by

\begin{eqnarray}
  \label{DistributionDef} n\left( \bi{r},t\right)
   & =&
\frac{\sum_{n=1}^{+\infty }a_{n}p_{n}\left( \bi{r}\right) g_{n}\left( t\right) }{\int
\sum_{n=1}^{+\infty }a_{n}p_{n}\left(
\bi{r}\right) g_{n}\left( t\right) \dif \bi{r}}    \nonumber \\
  & =& \frac{\sum_{n=1}^{+\infty }a_{n}p_{n}\left( \bi{r}\right)
g_{n}\left( t\right) }{\sum_{n=1}^{+\infty }a_{n}g_{n}\left( t\right) } \mbox{.}
\end{eqnarray}

The numerator is

\begin{eqnarray}
  \label{Auxiliary}
  \quad \sum_{n=1}^{+\infty }a_{n}p_{n}\left(
\bi{r}\right) g_{n}\left( t\right) =                   \nonumber \\
\quad = \sum_{n=1}^{m-1}a_{n}p_{n}\left( \bi{r}\right) g_{n}\left( t\right) + \frac{a_{nc}}{\alpha
_{nc}^{m-1}}p_{nc}\left( \bi{r}\right) \sum_{n=m}^{+\infty }\frac{\alpha
_{nc}^{n-1}t^{n-1}}{\left( n-1\right) !}\me ^{-t}          \nonumber \\
\quad = \sum_{n=1}^{m-1}a_{n}p_{n}\left( \bi{r}\right) g_{n}\left( t\right) +
                                                           \nonumber \\
\qquad + \frac{a_{nc}}{\alpha _{nc}^{m-1}}p_{nc}\left( \bi{r}\right) \left[ \sum_{n=1}^{+\infty
}\frac{\left( \alpha _{nc}t\right) ^{n-1}}{\left( n-1\right) !}\me
^{-t}-\sum_{n=1}^{m-1}\frac{\left( \alpha _{nc}t\right) ^{n-1}}{\left( n-1\right) !}\me
^{-t}\right]
  \mbox{.}
\end{eqnarray}

Using the incomplete Gamma function one finally has~(for an alternative formulation see the
Appendix)

\begin{eqnarray}
\sum_{n=1}^{m-1}a_{n}p_{n}\left( \bi{r}\right) g_{n}\left( t\right) + \frac{a_{nc}}{\alpha
_{nc}^{m-1}}p_{nc}\left( \bi{r}\right) \left[ \sum_{n=0}^{+\infty }\frac{\left( \alpha
_{nc}t\right) ^{n}}{n!}\me ^{-t}-\sum_{n=0}^{m-2}\frac{\left( \alpha _{nc}t\right)
^{n}}{n!}\me ^{-t}\right] =                                \nonumber \\
\quad = \sum_{n=1}^{m-1}a_{n}p_{n}\left( \bi{r}\right) g_{n}\left( t\right) +
                                                           \nonumber \\
\qquad + \frac{a_{nc}}{\alpha _{nc}^{m-1}}p_{nc}\left( \bi{r}\right) \left[ 1-\frac{\Gamma \left(
m-1,\alpha _{nc}t\right) }{\left( m-2\right) !}\right] \me ^{-\left( 1-\alpha _{nc}\right) t}
  \mbox{.}
\end{eqnarray}

The final form of the spatial distribution is thus

\begin{eqnarray}
  \label{Distribution:Main}
  n\left( \bi{r},t\right) =                            \nonumber \\
                                                           \nonumber \\
  \quad = \frac{\sum_{n=1}^{m-1}a_{n}p_{n}\left( \bi{r}\right) g_{n}\left(
t\right) + \frac{a_{nc}}{\alpha _{nc}^{m-1}}p_{nc}\left( \bi{r}\right) \left[ 1-\frac{\Gamma \left(
m-1,\alpha _{nc}t\right) }{\left( m-2\right) !}\right] \me ^{-\left( 1-\alpha _{nc}\right) t}}
{\sum_{n=1}^{m-1}a_{n}g_{n}\left( t\right) + \frac{a_{nc}}{\alpha _{nc}^{m-1}}\left[ 1-\frac{\Gamma
\left( m-1,\alpha _{nc}t\right) }{\left( m-2\right) !}\right] \me ^{-\left( 1-\alpha _{nc}\right)
t}}
  \mbox{.}
\end{eqnarray}

\subsection{The nonchanging distribution}\label{The nonchanging distribution}

The procedure used in~\sref{Ensemble relaxation} to define a \textit{nonchanging} spatial
distribution can now be checked in the last equation. From the zero time limit of the Gamma
function one has~$\lim_{x\rightarrow 0}\frac{\Gamma \left( m-1,x\right) }{\left( m-2\right) !}=1$
and only the first generation contributes. Therefore, at zero time, the excitation distribution is
the first generation distribution, as it should. However, for sufficiently long times so that all
the~$g_{n}\left( t\right)$ terms with~$n<m$ die out, $\lim_{t\rightarrow +\infty }n\left(
\bi{r},t\right) =p_{nc}\left( \bi{r}\right)$, the excitation distribution reducing to the
nonchanging value. Associated with this nonchanging distribution is a monoexponential decay kinetic
constant:

\begin{eqnarray}
\label{DecayFundamental}
\lim_{t\rightarrow +\infty }\rho \left(
t\right) =                                                 \nonumber \\
\quad = \lim_{t\rightarrow +\infty }\bigg\{ \sum_{n=1}^{m-1}q_{n}a_{n}g_{n}\left( t\right) +
\frac{q_{nc}a_{nc}}{\alpha _{nc}^{m-1}}\left[ 1-\frac{\Gamma \left( m-1,\alpha _{nc}t\right)
}{\left( m-2\right) !}\right] \me ^{-\left(
1-\alpha _{nc}\right) t}\bigg\}                           \nonumber \\
\quad = \frac{q_{nc}a_{nc}}{\alpha _{nc}^{m-1}}\me ^{-\left( 1-\alpha _{nc}\right) t}
  \mbox{.}
\end{eqnarray}

%%%%%%%%%%%%%%%%%%%%%%%%%%%%%%%%%%%%%%%%%%%%%%%%%%%%%%%%%%%%%%%%%%%%%%%%%%%%%%%%
%%% Steady-state conditions
%%%%%%%%%%%%%%%%%%%%%%%%%%%%%%%%%%%%%%%%%%%%%%%%%%%%%%%%%%%%%%%%%%%%%%%%%%%%%%%%

%
\section{Steady-state conditions}\label{Steady-state conditions}

The above expressions are strictly valid for a delta pulse excitation. However, under incoherent
conditions the decay for other initial excitation distributions is obtained from linear response
theory as the convolution of the excitation profile with the delta response
function~(e.g.~\cite{Berberan-Santos1992:ChemPhys.164.259} and references therein). Let us obtain
the system observables for a continuous excitation which constitutes a particularly important
limiting case in many practical conditions. We have then \textit{steady-state} or
\textit{stationary} conditions and use superscript~$SS$ to express it. The normalized spatial
distribution is obtained by time integrating both numerator and denominator
of~\eref{DistributionDef} giving

\begin{eqnarray}
\label{DistributionSS} n^{SS}\left( \bi{r}\right) & = & \frac{\sum_{n=1}^{+\infty }a_{n}p_{n}\left(
\bi{r}\right)
}{\sum_{n=1}^{+\infty }a_{n}}                              \nonumber \\
& = & \frac{\sum_{n=1}^{m-1}a_{n}p_{n}\left( \bi{r}\right) +\frac{a_{nc}}{1-\alpha
_{nc}}p_{nc}\left( \bi{r}\right) }{\sum_{n=1}^{m-1}a_{n}+\frac{a_{nc}}{1-\alpha _{nc}}}
  \mbox{,}
\end{eqnarray}

since the $g_{n}\left( t\right) $ are normalized. The overall emission intensity is just the
previously computed macroscopic emission yield~$\phi $ but now the most important quantity is the
spectral distribution. To obtain it, the decay should be resolved both in the optical frequency and
in the detection geometrical details. For the most important case of complete frequency
redistribution~\cite{Molisch1998:book} both spectra, absorption and emission, are the same and
therefore the decay is

\begin{eqnarray}
\rho ^{\Omega } \left( x,t\right) =\sum_{n=1}^{+\infty }\Phi \left( x\right) q_{n}^{\Omega }\left(
x\right) a_{n}g_{n}\left( t\right)
  \mbox{.}
\end{eqnarray}

The decay at each frequency depends upon the intrinsic spectrum~$\Phi \left( x\right)$  and on the
mean escape probability in the detection direction~$\Omega$,

\begin{eqnarray}
\label{EscapeOmega} q_{n}^{\Omega }\left( x\right) =\int_{\Omega }\! \int_{V}\! \me ^{-\Phi \left(
x\right) r}p_{n}\left( \bi{r}\right) d\bi{r}\dif S
  \mbox{,}
\end{eqnarray}

where the Beer-Lambert escape~(survival) probability is weighted in both spatial distribution
inside volume~$V$ and over the surface~$S$ facing detection optics.~$r$ should be the optical
distance between emission coordinate~$\bi{r} $ and the surface point facing detection.

Now the steady-state spectra is obtained by time integrating this, giving

\begin{eqnarray}
I^{SS, \Omega }\left( x\right) =\sum_{n=1}^{+\infty }\Phi \left( x\right) q_{n}^{\Omega }\left(
x\right) a_{n}
  \mbox{.}
\end{eqnarray}

The trapping distortion of emission spectra is more informative if one corrects for emission
intensity scale factors and it is thus better to have the normalized spectral distribution,

\begin{eqnarray}
\label{ISS}
I^{SS, \Omega }\left( x\right) =\frac{\sum_{n=1}^{m-1}q_{n}^{\Omega }\left( x\right)
a_{n}+\frac{q_{nc}^{\Omega }\left( x\right) a_{nc}}{1-\alpha _{nc}}}{\int_{-\infty }^{+\infty
}\left[ \sum_{n=1}^{m-1}q_{n}^{\Omega }\left( x\right) a_{n}+\frac{q_{nc}^{\Omega }\left( x\right)
a_{nc}}{1-\alpha _{nc}}\right] \Phi \left( x\right) dx}\Phi \left( x\right)
  \mbox{.}
\end{eqnarray}

The numerator gives the trapping dependent spectral distortion. At this point there must be
stressed out the paramount importance of the~$x$-dependent escape probability in weighting each
generation contribution to the observed spectra, this being due to the strong nonlinear character
of~\eref{EscapeOmega}.

%%%%%%%%%%%%%%%%%%%%%%%%%%%%%%%%%%%%%%%%%%%%%%%%%%%%%%%%%%%%%%%%%%%%%%%%%%%%%%%%
%%% Fundamental mode
%%%%%%%%%%%%%%%%%%%%%%%%%%%%%%%%%%%%%%%%%%%%%%%%%%%%%%%%%%%%%%%%%%%%%%%%%%%%%%%%

%
\section{Fundamental mode}\label{Fundamental mode}

A simple stochastic formulation of the multiple scattering representation~(MSR) solution to the
classical Holstein-Biberman equation is presented for a broad audience. This is made using the
definition of several generations of excited species, according with the number of previous
emission-reabsorption events. The contribution of these generations is divided into two groups, one
for generations whose normalized spatial distribution changes and the other for
\textit{nonchanging} generations. It was shown that this last group gives rise to a monoexponential
term in the relaxation dynamics of the ensemble. Comparing the~MSR solution to the alternative
Holstein exponential expansion, this monoexponential term can be identified with Holstein's well
known \textit{fundamental} relaxation mode in \eref{Holstein-Solution}; the \textit{slowest}
decaying mode (the one corresponding to the smallest eigenvalue and the only one with only positive
values for the corresponding spatial profile). This provides a clear connection between MSR and the
older practice of using Holstein mode expansion.

The Holstein ansatz is still the most commonly used model to quantify atomic trapping. However,
given the notorious difficulties in obtaining all the eigenmodes/decay parameters for realistic
spectral and geometrical conditions, trapping is usually reduced to the consideration of the
slowest decaying~(fundamental) mode. Moreover, for this mode and usually in conditions where
trapping is not the main focus of research, this is further simplified by the use Holstein's
asymptotic expansion~\cite{Molisch1998:book}. This is strictly valid only in the limit of very high
opacities and for ideal~1D geometries difficult to fulfil in realistic setups. Given the
accumulated body of data in the literature for this practice, a critical assessment of the
conditions in which this is judged to be adequate enough is useful.

There is a qualitative important difference between time-resolved quantities~(decay and spatial
distribution) on the one side, and overall relaxation parameters~(macroscopic ensemble emission
yield and mean scaled lifetime) or steady-state~(spectra and spatial distribution) quantities on
the other. By waiting long enough for excitation to relax into the~\textit{nonchanging}
distribution, the decay and overall spatial distribution will converge to a monoexponential mode.
However the mean reemission yield~\textendash ~\eref{phi}~\textendash, and the steady-state
spatial~\textendash~\eref{DistributionSS}~\textendash~ and, spectral
distributions~\textendash~\eref{ISS}~\textendash~ will always have contributions from all the
generations. This justify two experimental well know procedures to increase the applicability of
Holstein's fundamental mode:~(i)~in time resolved setups, wait long enough until relaxation is
judged to be sufficiently monoexponential by fitting the data, and~(ii)~in all experiments, design
the set up for primordial excitation to mimic as best as possible the fundamental mode spatial
distribution~(symmetrical in cell and more or less constant). This tailoring of the first
generation excitation to the fundamental distribution is usually made either with electron impact
excitation or with strongly detuned photoexcitation. However, these practices can only alleviate
the limitations imposed by the sole use of the fundamental mode to an acceptable degree but cannot
circumvent in full the fact that the overall relaxation and steady-state will have contributions
from all of the Holstein's modes~(this is specially important for the steady-state emission spectra
given the strongly nonlinear dependence of detection probabilities in~\eref{EscapeOmega} on the
spatial distribution functions).

The previous equations can be used to trivially obtain the relative contribution of the nonchanging
mode to the sought for parameters. These contributions are,

\begin{eqnarray}
  \Theta _{nc,\Phi }
  =\frac{\frac{q_{nc}a_{nc}}{1-\alpha _{nc}}}{\Phi }
  \mbox{,}
\end{eqnarray}

\begin{eqnarray}
\Theta _{nc,\tau }=\frac{\frac{q_{nc}a_{nc}}{\left( 1-\alpha _{nc}\right) ^{2}}\left[ m\left(
1-\alpha _{nc}\right) +\alpha _{nc}\right] }{\sum_{n=1}^{m-1}nq_{n}a_{n}+\frac{q_{nc}a_{nc}}{\left(
1-\alpha _{nc}\right) ^{2}}\left[ m\left( 1-\alpha _{nc}\right) +\alpha _{nc}\right] }
  \mbox{,}
\end{eqnarray}

\begin{eqnarray}
\Theta _{nc,n^{SS}\left( r\right) }=\frac{\frac{a_{nc}}{1-\alpha _{nc}}p_{nc}\left( r\right)
}{\sum_{n=1}^{m-1}a_{n}+\frac{a_{nc}}{1-\alpha _{nc}}}
  \mbox{,}
\end{eqnarray}

and

\begin{eqnarray}
\Theta _{nc,I^{SS}\left( x\right) }=\frac{\frac{q_{nc}^{\Omega }\left( x\right) a_{nc}}{1-\alpha
_{nc}}}{\int_{-\infty }^{+\infty }\left[ \sum_{n=1}^{m-1}q_{n}^{\Omega }\left( x\right)
a_{n}+\frac{q_{nc}^{\Omega }\left( x\right) a_{nc}}{1-\alpha _{nc}}\right] \Phi \left( x\right)
dx}\Phi \left( x\right)
  \mbox{,}
\end{eqnarray}

and should always be checked to see if the use of the fundamental mode alone does not constitute a
severe approximation in any given experimental situation.

%%%%%%%%%%%%%%%%%%%%%%%%%%%%%%%%%%%%%%%%%%%%%%%%%%%%%%%%%%%%%%%%%%%%%%%%%%%%%%%%
%%% Conclusions
%%%%%%%%%%%%%%%%%%%%%%%%%%%%%%%%%%%%%%%%%%%%%%%%%%%%%%%%%%%%%%%%%%%%%%%%%%%%%%%%

%
\section{Conclusions}\label{Conclusions}

The solution of the classical trapping problem is outlined for both Holstein's original ansatz and
the alternative multiple scattering representation~(MSR). Both approaches are equivalent and the
physical interpretation of the Holstein fundamental mode as the result of a nonchanging excitation
spatial distribution is discussed at length. Given the common practice of quantifying trapping by
the fundamental mode alone, particular attention is paid to show under which conditions this
procedure does not impose severe limitations~(time resolved quantities with primordial excitation
mimicking fundamental mode distribution) and when this is not justified~(overall relaxation
parameters and steady state quantities, specially spectra). The quantification of the fundamental
mode contribution to trapping dependent observables is also made.

The multiple scattering representation is given a clear stochastic formulation stressing physical
insight. The main limitations of the model are linear nonsaturating system response and incoherent
trapping with complete frequency redistribution in lab reference frame. No further limitations on
either spectral distributions, opacity limits of applicability or geometrical details are
assumed~(a minor trivial modification is necessary for molecular spectra in condensed phase since
in this case absorption and emission are different even in complete frequency redistribution
conditions).

The advantages of the MSR over Holstein's modes will be illustrated with a simple case study in the
following instalment of this work in which a simple Markov random walk like algorithm will be used
to quantify trapping for Doppler, Lorentz or Voigt complete frequency redistribution two-level
atomic models.

%%%%%%%%%%%%%%%%%%%%%%%%%%%%%%%%%%%%%%%%%%%%%%%%%%%%%%%%%%%%%%%%%%%%%%%%%%%%%%%%
%%% Acknowledgments
%%%%%%%%%%%%%%%%%%%%%%%%%%%%%%%%%%%%%%%%%%%%%%%%%%%%%%%%%%%%%%%%%%%%%%%%%%%%%%%%

\ack

This work was supported by Funda\c{c}\~{a}o para a Ci\^{e}ncia e Tecnologia~(Portugal) and
Universidade do Minho~(Portugal) within project~REEQ/433/EEI/2005. It also used computational
facilities bought under project~POCTI/CTM/41574/2001, funded by FCT and the European Community
Fund~FEDER. A.R.~Alves-Pereira acknowledges~FCT funding under the reference~SFRH/BD/4727/2001.
E.~Nunes-Pereira acknowledges the critical reading of the manuscript by M.~Besley~(Centro de
F\'{i}sica, Universidade do Minho).

%%%%%%%%%%%%%%%%%%%%%%%%%%%%%%%%%%%%%%%%%%%%%%%%%%%%%%%%%%%%%%%%%%%%%%%%%%%%%%%%
%%% Acknowledgments
%%%%%%%%%%%%%%%%%%%%%%%%%%%%%%%%%%%%%%%%%%%%%%%%%%%%%%%%%%%%%%%%%%%%%%%%%%%%%%%%

\appendix
\section*{Appendix: Alternative form for the spatial distribution}\label{Appendix}
\setcounter{section}{1}

If the generation number for the nonchanging distribution is very high then the computer evaluation
of terms with the incomplete Gamma can become troublesome since in this case it is a delicate
balance of two infinities. To avoid this~\eref{Auxiliary} can be rearranged instead into

\begin{eqnarray}
\sum_{n=1}^{m-1}a_{n}p_{n}\left( \bi{r}\right) g_{n}\left( t\right) + \frac{a_{nc}}{\alpha
_{nc}^{m-1}}p_{nc}\left( \bi{r}\right) \left[ \sum_{n=0}^{+\infty }\frac{\left( \alpha
_{nc}t\right) ^{n}}{n!}\me ^{-t}-\sum_{n=1}^{m-1}\frac{\left( \alpha _{nc}t\right) ^{n-1}}{\left(
n-1\right) !}\me ^{-t}\right] =
                                                           \nonumber \\
\quad = \sum_{n=1}^{m-1}a_{n}p_{n}\left( \bi{r}\right) g_{n}\left( t\right) -\frac{a_{nc}}{\alpha
_{nc}^{m-1}}p_{nc}\left( \bi{r}\right) \sum_{n=1}^{m-1}\alpha _{nc}{}^{n-1}g_{nc}\left( t\right) +
                                                           \nonumber \\
\qquad + \frac{a_{nc}}{\alpha _{nc}^{m-1}}p_{nc}\left( \bi{r}\right) \sum_{n=0}^{+\infty
}\frac{\left( \alpha_{nc}t\right) ^{n}}{n!}\me ^{-t}
                                                           \nonumber \\
\quad = \left[ \sum_{n=1}^{m-1}a_{n}p_{n}\left( \bi{r}\right) g_{n}\left( t\right)
-\frac{a_{nc}}{\alpha _{nc}^{m-1}}p_{nc}\left( \bi{r}\right) \sum_{n=1}^{m-1}\alpha
_{nc}{}^{n-1}g_{nc}\left( t\right) \right] +
                                                           \nonumber \\
\qquad + \frac{a_{nc}}{\alpha _{nc}^{m-1}}p_{nc}\left( \bi{r}\right) \me ^{-\left( 1-\alpha
_{nc}\right) t}
  \mbox{,}
\end{eqnarray}

which gives finally

\begin{eqnarray}
\label{Distribution:Appendix}
n\left( \bi{r},t\right) =                                  \nonumber \\
                                                           \nonumber \\
\quad = \frac{
\begin{array}{r}
\left[\sum_{n=1}^{m-1}a_{n}p_{n}\left( \bi{r}\right) g_{n}\left(
t\right) -\frac{a_{nc}}{\alpha _{nc}^{m-1}}p_{nc}\left( \bi{r}\right) \sum_{n=1}^{m-1}\alpha
_{nc}{}^{n-1}g_{nc}\left( t\right)
\right] + \\
+ \frac{a_{nc}}{\alpha _{nc}^{m-1}}p_{nc}\left( \bi{r}\right) \me ^{-\left( 1-\alpha _{nc}\right)
t}
\end{array}
}{
\begin{array}{r}
\left[ \sum_{n=1}^{m-1}a_{n}g_{n}\left( t\right) -\frac{a_{nc}}{\alpha
_{nc}^{m-1}}\sum_{n=1}^{m-1}\alpha _{nc}{}^{n-1}g_{nc}\left( t\right) \right] + \\
+ \frac{a_{nc}}{\alpha _{nc}^{m-1}}p_{nc}\left( \bi{r}\right) \me ^{-\left( 1-\alpha_{nc}\right)t}
\end{array}
} \mbox{,}
\end{eqnarray}

an alternative form for the spatial distribution given by \eref{Distribution:Main}, less compact
but also less prone to numerical artifacts.

%%%%%%%%%%%%%%%%%%%%%%%%%%%%%%%%%%%%%%%%%%%%%%%%%%%%%%%%%%%%%%%%%%%%%%%%%%%%%%%%
%%% Vertical space
%%%%%%%%%%%%%%%%%%%%%%%%%%%%%%%%%%%%%%%%%%%%%%%%%%%%%%%%%%%%%%%%%%%%%%%%%%%%%%%%

\vspace{1cm}

%%%%%%%%%%%%%%%%%%%%%%%%%%%%%%%%%%%%%%%%%%%%%%%%%%%%%%%%%%%%%%%%%%%%%%%%%%%%%%%%
%%% Comands for BibTeX
%%%%%%%%%%%%%%%%%%%%%%%%%%%%%%%%%%%%%%%%%%%%%%%%%%%%%%%%%%%%%%%%%%%%%%%%%%%%%%%%

%\bibliography{/Eduardo-J-Nunes-Pereira/Docs/Papers/Bibliography/Eduardo-J-Nunes-Pereira}

\begin{thebibliography}{99}

\bibitem{Mihalas1978:book}
D.~Mihalas, {\em Stellar Atmospheres}, 2$^{nd}$~Ed. (Freeman, San Francisco, 1978).

\bibitem{Molisch1998:book}
A.~F. Molisch and B.~P. Oehry, {\em Radiation Trapping in Atomic Vapours} (Oxford, Oxford, 1998).

\bibitem{Thomas1999:book}
G.~E.~Thomas and K.~Stamnes, {\em Radiative Transfer in the Atmosphere and Ocean} (Cambridge,
Cambridge, 1999).

\bibitem{Berberan-Santos1999:chapter}
M.~N.~Berberan-Santos, E.~Pereira, and J.~M.~G.~Martinho, {\em Dynamics of radiative transport}, in
D.~L. Andrews and A.~A. Demidov, editors, {\em Resonance Energy Transfer} (John Wiley \& Sons,
Chichester, 1999), p.~108.

\bibitem{Modest2003:book}
M.~F.~Modest, {\em Radiative Heat Transfer}, 2$^{nd}$~Ed. (Academic Press, San Diego, 2003).

\bibitem{Bardou2002:book}
F.~Bardou, J.~P.~Bouchaud, A.~Aspect, and C.~Cohen-Tannoudji, {\em L\'{e}vy Statistics and Laser
Cooling} (Cambridge, Cambridge, 2002).

\bibitem{Holstein1947:PhysRev.72.1212}
T.~Holstein, {\em Phys. Rev.}, \textbf{72}, 1212 (1947).

\bibitem{Biberman1947:ZhFiz.17.416}
L.~M.~Biberman, {\em Zh. Eksperim. i Teor. Fiz.}, \textbf{17}, 416 (1947).

\bibitem{Nunes-Pereira2004:PhysRevLett.93.120201}
E.~Pereira, J.~M.~G.~Martinho, and M.~N.~Berberan-Santos, {\em Phys. Rev. Lett.}, \textbf{93},
120201 (2004).

\bibitem{Metzler-Anomalous-2000:PhysRep.399.1&Metzler-Anomalous-2004:JPhysA.37.R161&SuperdiffusionI:JCP.125.174308}
R.~Metzler and J.~Klafter, {\em Phys. Rep.}, \textbf{399}, 1 (2000); {\em J. Phys. A-Math. Gen.},
\textbf{37}, R161 (2004); M.~N.~Berberan-Santos, E.~J. Nunes-Pereira, and J.~M.~G.~Martinho. {\em
J. Chem. Phys.}, \textbf{125}, 174308 (2006).

\bibitem{Lister2004:RMP.76.541&Proud1986:book&Waymouth1971:book}
G.~G.~Lister, J.~E.~Lawler, W.~P.~Lapatovich, and V.~A.~Godyak, {\em Rev. Mod. Phys.}, \textbf{76},
541 (2004); J.~M.~Proud and L.~H.~Luessen, editors, {\em Radiative Processes in Discharge Plasmas}
(NATO ASI Series, Series B Physics, Plenum Press, New York, 1986); J.~F.~Waymouth. {\em Electric
Discharge Lamps} (MIT Press, Cambridge Massachusets, 1971).

\bibitem{Kushner2004:JPhysD.37.R161}
K.~Rajaraman and M.~J.~Kushner, {\em J. Phys. D-Appl. Phys.}, \textbf{37}, 1780 (2004).

\bibitem{KushnerPDPsB:JApplPhys.85.3460&KushnerPDPsC:JApplPhys.85.3470&KushnerPDPsA:JApplPhys.87.2700}
S.~Rauf and M.~J. Kushner, {\em J. Appl. Phys.}, \textbf{85}, 3460 (1999); {\em J. Appl. Phys.},
\textbf{85}, 3470 (1999); T.~van~der~Straaten and M.~J.~Kushner, {\em J. Appl. Phys.}, \textbf{87},
2700 (2000).

\bibitem{Falecki1989:ZPhysD.14.111}
W.~Falecki, W.~Hartmann, P.~and Wiorkowski, {\em Z. Phys. D}, \textbf{14}, 111 (1989).

\bibitem{Martinho1989:JCP.90.53}
J.~M.~G.~Martinho, A.~L.~Ma\c{c}anita, and M.~N.~Berberan-Santos, {\em J. Chem. Phys.},
\textbf{90}, 53 (1989).

\bibitem{Lai1993:ZPhysD.27.223&Lai1993:OpticsComm.99.316}
R.~Lai, S.~L.~Liu, and X.~X.~Ma, {\em Z. Phys. D}, \textbf{27}, 223, 1993; {\em Optics Comm.},
\textbf{99}, 316 (1993).

\bibitem{Berberan-Santos1992:ChemPhys.164.259}
M.~N.~Berberan-Santos and J.~M.~G.~Martinho, {\em Chem. Phys.}, \textbf{164}, 259 (1992).

\end{thebibliography}
%\bibliographystyle{unsrt}

%%%%%%%%%%%%%%%%%%%%%%%%%%%%%%%%%%%%%%%%%%%%%%%%%%%%%%%%%%%%%%%%%%%%%%%%%%%%%%%%
%%% end Comands for BibTeX
%%%%%%%%%%%%%%%%%%%%%%%%%%%%%%%%%%%%%%%%%%%%%%%%%%%%%%%%%%%%%%%%%%%%%%%%%%%%%%%%

%\end{linenumbers}

\end{document}